\documentclass[sigconf,anonymous=False]{acmart}

\AtBeginDocument{%
  }

\copyrightyear{2025}
\acmYear{2025}
\setcopyright{acmlicensed}
\acmConference[WSDM '25]{Proceedings of the Eighteenth ACM International Conference on Web Search and Data Mining}{March 10--14, 2025}{Hannover, Germany}
\acmBooktitle{Proceedings of the Eighteenth ACM International Conference on Web Search and Data Mining (WSDM '25), March 10--14, 2025, Hannover, Germany}
\acmDOI{10.1145/3701551.3703572}
\acmISBN{979-8-4007-1329-3/25/03}

\usepackage{enumitem}
\usepackage{color}
\usepackage{soul}
\usepackage{balance}
\usepackage[normalem]{ulem}
\usepackage{multirow}

\begin{document}

\title{DLCRec: A Novel Approach for Managing Diversity in LLM-Based Recommender Systems}


\author{Jiaju Chen}
\email{cjj01@mail.ustc.edu.cn}
\affiliation{%
  \institution{University of Science and Technology of China}
  \city{Hefei}
  \country{China}
}

\author{Chongming Gao}
\authornote{Corresponding author.}
\email{chongming.gao@gmail.com}
\affiliation{%
  \institution{University of Science and Technology of China}
  \city{Hefei}
  \country{China}
}

\author{Shuai Yuan}
\email{syuanaf@connect.ust.hk}
\affiliation{%
  \institution{Hong Kong University of Science and Technology}
  \city{Hong Kong}
  \country{China}
}

\author{Shuchang Liu}
\email{1183965286@qq.com}
\affiliation{%
  \institution{Independent}
  \city{Beijing}
  \country{China}
}

\author{Qingpeng Cai}
\email{cqpcurry@gmail.com}
\affiliation{%
  \institution{Independent}
  \city{Beijing}
  \country{China}
}

\author{Peng Jiang}
\email{jp2006@139.com}
\affiliation{%
  \institution{Independent}
  \city{Beijing}
  \country{China}
}

\renewcommand{\shortauthors}{Jiaju Chen et al.}

\begin{abstract}
The integration of Large Language Models (LLMs) into recommender systems has led to substantial performance improvements. However, this often comes at the cost of diminished recommendation diversity, which can negatively impact user satisfaction. To address this issue, controllable recommendation has emerged as a promising approach, allowing users to specify their preferences and receive recommendations that meet their diverse needs. Despite its potential, existing controllable recommender systems frequently rely on simplistic mechanisms, such as a single prompt, to regulate diversity—an approach that falls short of capturing the full complexity of user preferences. In response to these limitations, we propose DLCRec, a novel framework designed to enable fine-grained control over diversity in LLM-based recommendations. Unlike traditional methods, DLCRec adopts a well-designed task decomposition strategy, breaking down the recommendation process into three sequential sub-tasks: genre prediction, genre filling, and item prediction. These sub-tasks are trained independently and inferred sequentially according to user-defined control numbers, ensuring more precise control over diversity. Furthermore, the scarcity and uneven distribution of diversity-related user behavior data pose significant challenges for fine-tuning. To overcome these obstacles, we introduce two data augmentation techniques that enhance the model's robustness to noisy and out-of-distribution data. These techniques expose the model to a broader range of patterns, improving its adaptability in generating recommendations with varying levels of diversity. Our extensive empirical evaluation demonstrates that DLCRec not only provides precise control over diversity but also outperforms state-of-the-art baselines across multiple recommendation scenarios.
\end{abstract}



\maketitle

\section{Introduction}
\begin{figure}[tb!]
\setlength{\belowcaptionskip}{-0.5cm}
\centering
\includegraphics[scale=0.4]{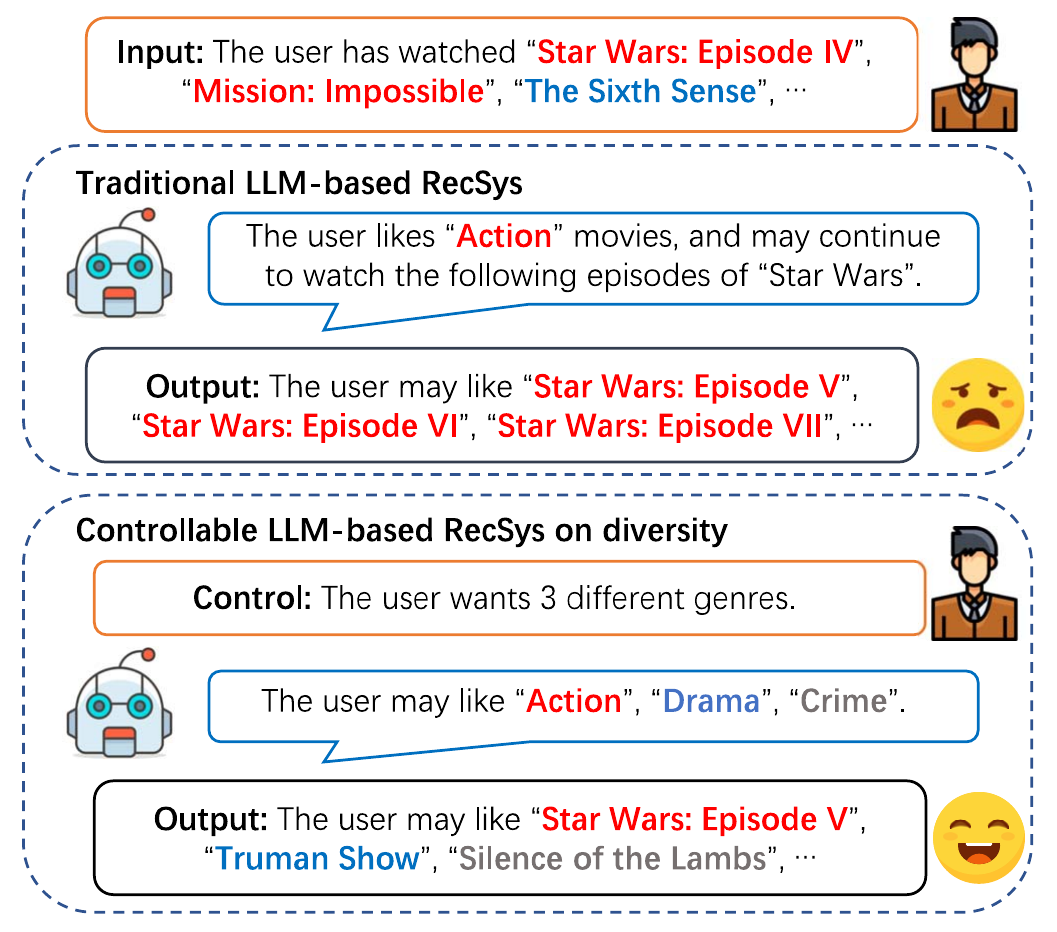}
\caption{Illustration of the difference between traditional and controllable LLM-based recommender systems.}
\label{fig:intro}
\end{figure}

The advent of Large Language Models (LLMs) has significantly augmented the capabilities of recommender systems, enabling few-shot recommendations that harness semantic understanding ~\cite{bao2023tallrec,bao2023bi}. However, a major limitation of LLM-based recommenders is their tendency to produce homogeneous results that closely mirror users' historical interactions ~\cite{lu2024aligning, bao2024decoding}, as is shown in Figure \ref{fig:intro}. This homogeneity can lead to issues such as echo chambers and filter bubbles ~\cite{gao2023cirs,gao2023alleviating,wang2022user}, ultimately compromising user satisfaction. In response, the field has shifted from solely prioritizing accuracy to considering a broader range of recommendation quality metrics, including diversity ~\cite{petrov2024aligning, lu2024aligning}. Moreover, empowering users to control their recommended results is increasingly seen as key to fostering a more inclusive and engaging user experience. Providing users with control over the diversity of their recommendations allows recommender systems to better align with individual preferences, facilitate interest discovery ~\cite{chen2024treatment}, and mitigate the risks associated with algorithmic monocultures ~\cite{gao2023alleviating}.



Given the extensive semantic information embedded within LLMs, \citet{liang2024taxonomy} highlights the intrinsic capacity of these models to comprehend and potentially manage the diversity of recommendations according to users' needs. However, directly applying these models to controllable recommendation tasks with varying diversity requirements presents significant challenges, primarily due to the absence of explicit training and alignment tailored to such tasks.
While existing LLM-based controllable recommender systems often rely on a single prompt to regulate diversity~\cite{gao2024llm}, our findings suggest that this approach may be overly simplistic. Traditional fine-tuning methods~\cite{lu2024aligning} and Chain-of-Thought (CoT) prompting techniques~\cite{wei2022chain} may not provide the nuanced control necessary for effectively managing diversity in recommendations \cite{shi2024sigir}. This challenge is exacerbated by the sparsity and uneven distribution of diversity-related needs extracted from user behavior~\cite{lu2024aligning, petrov2024aligning,gao2021advances}. As a result, LLMs struggle to develop a sophisticated understanding of diversity control from these limited examples.

To address these challenges, we propose a novel approach, termed as \textbf{D}iversity-oriented \textbf{L}LM-based \textbf{C}ontrollable \textbf{Rec}ommender (DLCRec), which enables precise control over diversity in LLM-based recommendations. DLCRec generates recommendations with varying levels of diversity—defined as the coverage of multiple categories within a recommendation list—and optimizes the recommender system toward this objective.
Our approach is built on a fine-grained task decomposition strategy, which guides the LLM to fully leverage its capabilities for diverse recommendations. Specifically, we break down the complex recommendation task into three sequential sub-tasks: (1) predicting the genres of future recommendations, (2) filling in the genres, and (3) predicting the future items. We leverage few-shot samples to construct task-specific data for each sub-task's fine-tuning. In the training phase, we train the three sub-tasks separately, while in the inference phase, DLCRec processes user-specified control numbers—specifically, the desired numbers of genres in the recommendation lists—and executes the sub-tasks in sequence to generate personalized recommendations that match users' diversity preferences. To facilitate a seamless connection between the latter two tasks, we employ a concise and fixed output format to constrain the LLMs' output. 
%

During the inference phase, control targets may be unseen or fall outside the distribution of the training data, potentially compromising the model's ability to generalize to novel controlled recommendation results. To address this challenge and mitigate the sparsity and skewness of diversity signals in the data, we introduce two data augmentation methods tailored for the genre filling and item prediction tasks. 
First, we inject noise into the training samples to simulate errors in preceding tasks, thereby enhancing the model's robustness to control targets that may be incorrectly predicted by earlier stages. Second, we modify the distributions of control numbers to include underrepresented or extreme cases to better equip DLCRec to handle scenarios that are rarely encountered during training. By integrating these augmented datasets with the original training data, we significantly enhance the model's robustness and adaptability in genre-controlled item generation.

We comprehensively evaluate the effectiveness of our proposed method through extensive experiments on two real-world datasets, comparing it against a range of competitive baseline approaches. The empirical results demonstrate that our method achieves precise control of diversity, with only a marginal sacrifice in accuracy, thereby showcasing its promising potential for real-world applications where diversity is a critical consideration. We release our code and data at \url{https://github.com/Jiaju-Chen/DLCRec}.

The main contributions of our work are summarized as follows:
\begin{itemize}[leftmargin=*]
\item We propose a novel paradigm for diversity-oriented controllable recommendations in LLMs, which enables fine-grained control over the diversity of recommended items.
\item We introduce DLCRec, a new framework for controllable recommendations on diversity, which decomposes the complex controllable recommendation task into manageable sub-tasks and leverages data augmentation techniques to improve the model's robustness and adaptability to diverse control numbers.
\item We conduct a comprehensive empirical evaluation of DLCRec, demonstrating its effectiveness in providing precise control over diversity and outperforming state-of-the-art baselines in various recommendation scenarios.
\end{itemize}

\begin{figure*}[!t]
\setlength{\abovecaptionskip}{0cm}
\setlength{\belowcaptionskip}{-0.1cm}
    \centering
    \includegraphics[width=0.75\linewidth]{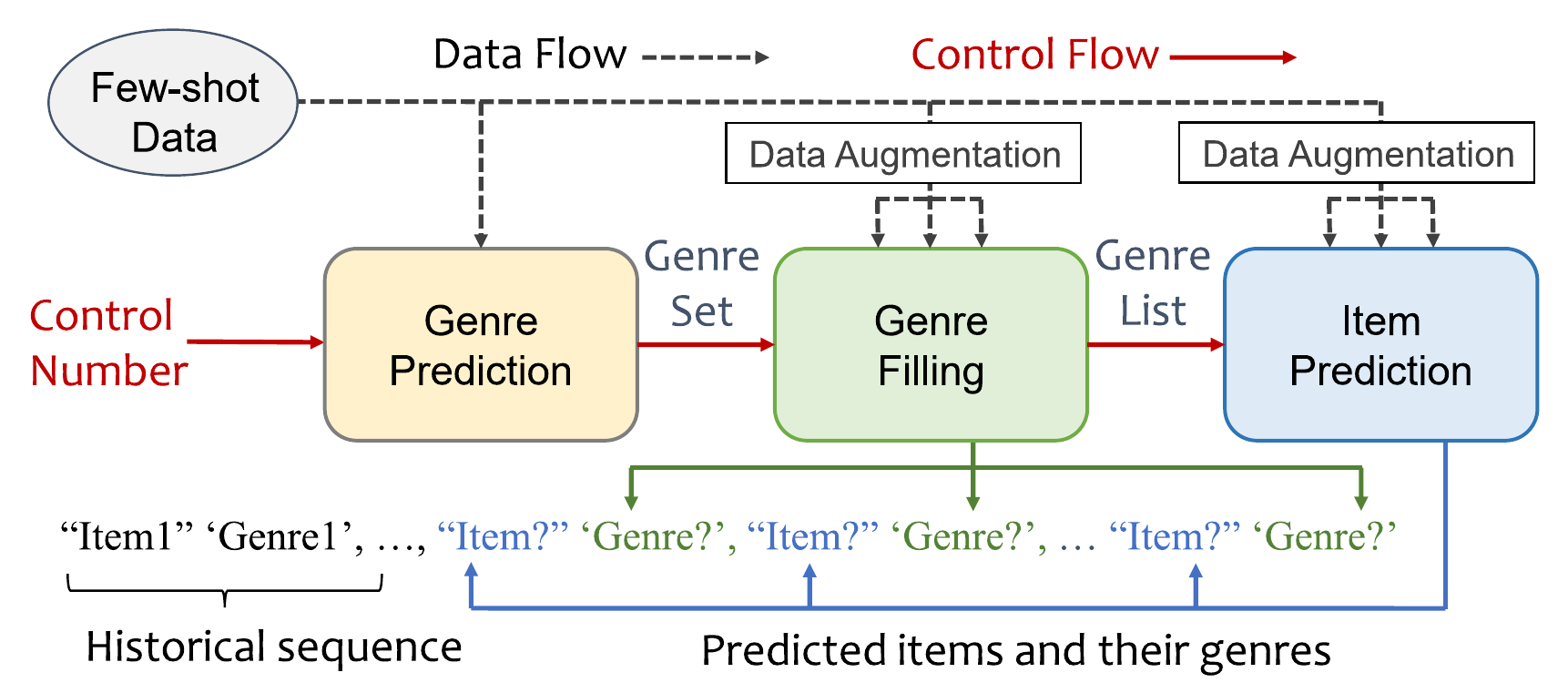}
    \caption{Overview of the DLCRec framework. DLCRec decomposes the recommendation task into three sub-tasks: Genre Predicting, Genre Filling, and Item Predicting. The framework consists of two complementary components: the training framework (black lines) and the control framework (read lines). In the training framework, we leverage few-shot data and data augmentation to train each sub-task independently. In contrast, the control framework enables explicit control over the diversity of the final recommendation list by propagating the control number throughout the three sub-tasks. }
    \label{fig:framework}
\end{figure*}

\section{Related Work}
In this section, we provide a brief overview of LLM-based recommender systems and explore existing studies that leverage LLMs for controllable recommendations.
\subsection{LLMs for Recommendation}
The emergence of powerful LLMs such as GPT4~\cite{achiam2023gpt} and Llama3~\cite{dubey2024llama} has sparked a growing interest in integrating LLMs into recommender systems \cite{lin2023can,wu2023survey,zhao2023recommender}. A common approach is to fine-tune LLMs for specific recommendation tasks~\cite{wang2023recmind,bao2023bi,lin2023multi,yang2023palr, gao2024sprec, liao2024rosepo}. Leveraging their exceptional semantic understanding capabilities, LLM-based recommenders have demonstrated superior performance over traditional ID-based models, particularly in cold-start or few-shot learning scenarios~\cite{bao2023bi,liang2024taxonomy}. However, these fine-tuning approaches often rely on explicit signals that are relatively easy to learn, and their limitations become apparent when dealing with more complex signals, such as diversity. In contrast, our work proposes a task decomposition approach, breaking down the complex diversity-control recommendation problem into several manageable tasks, better suited for the fine-tuning framework.
Other studies have focused on augmenting LLM-based recommenders with collaborative filtering information~\cite{zhang2023collm,liao2023llara}, such as item embeddings from traditional models, which enhances their performance in areas with dense data. While these approaches have shown promise, our work takes a different direction, exploring the inherent capacity of LLMs to control diversity in recommendations. By doing so, we aim to unlock the full potential of LLMs in recommender systems and provide a more nuanced understanding of their capabilities.

\subsection{LLMs for Controllable Recommendation}
Current LLM-based controllable recommendation approaches can be broadly categorized into two groups: instruction tuning~\cite{gao2024llm, wang2023recmind, geng2022recommendation, zhang2023recommendation} and reinforcement learning fine-tuning (RLFT)~\cite{lu2024aligning,sharma2024optimizing,petrov2024aligning}.
Instruction tuning involves fine-tuning LLMs on various tasks to enable generalization across diverse user needs. For instance, P5~\cite{geng2022recommendation} and InstructRec~\cite{zhang2023recommendation} employ multiple recommendation tasks to cater to different user requirements. \citet{lu2024aligning} proposed prompt-based control to modulate the proportion of a single category. However, ensuring diversity while maintaining high accuracy is a more complex challenge that may not be effectively addressed by single-task instructions, as it often requires balancing multiple genres.
On the other hand, RLFT methods ~\cite{lu2024aligning,petrov2024aligning} carefully design reward signals to refine the LLM's ability to follow instructions after supervised learning. Nevertheless, RLFT approaches are often data-intensive and suffer from convergence issues. In contrast, our approach leverages data augmentation to enhance model robustness to rare and noisy controls, offering an efficient and effective solution for scenarios with sparse data.

\begin{table*}[!t]
    \caption{Illustration of the three sub-tasks in the DLCRec framework: Genre Predicting (GP), Genre Filling (GF), and Item Predicting (IP). The highlighted regions (in yellow) indicate the control points for modifying the input prompts in the control framework. (a) Task GP predicts future genres based on historical interaction trails. In the control framework, we adjust the desired number of genres. (b) Task GF fills in future genres given target genres and historical interaction trails. In the control framework, we modify the target genres. (c) Task IP predicts future items based on predicted future genres and historical interaction trails. In the control framework, we adjust the predicted future genres to control the recommendation output.}
  \label{tab:sub-tasks}
  \centering
  \resizebox{\textwidth}{!}{
  \begin{tabular}{|p{0.33\textwidth}|p{0.33\textwidth}|p{0.33\textwidth}|}
    \hline
    
    \textbf{(a) Genre predicting (GP)} & \textbf{(b) Genre filling (GF)} & \textbf{(c) Item predicting (IP)} \\
    \hline
    \textbf{Instruction:} & \textbf{Instruction:} & \textbf{Instruction:} \\
    Given a list of movies and their corresponding genres the user has watched before, \sethlcolor{yellow}\hl{please provide the 3 most likely genres in the future recommendation list.} Output the genres only, without movie names, explanations, or numbers. The output format is ``Genre1, Genre2, $\cdots$, Genre3''.
    & Below is a user's interaction trail of movies he likes. Each movie is in double quotes ``'', followed by its genre in single quotes `'. \sethlcolor{yellow}\hl{Your task is to fill in the genre placeholders represented by ``?'' with the following genres: [Action, Comedy, Drama].} The ``\_'' represents placeholder tokens that you should not consider. The output should maintain the same format as the input.
    & Below is a user's interaction trail of movies he likes. Each movie is in double quotes ``'', followed by its genre in single quotes `'. Your task is to fill in the movie placeholders represented by ``?''. The ``\_'' represents placeholder tokens that you should not consider. The output should maintain the same format as the input.
    \\
    \hline

    \textbf{Input:} & \textbf{Input:} & \textbf{Input:} \\
    The user has watched the following movies with their corresponding genres in ``()'' before: ``Star Wars: Episode V - The Empire Strikes Back (1980)'' (Action), ``Mission: Impossible (1996)'' (Action), $\cdots$, ``Stargate (1994)'' (Action).
    & ``Star Wars: Episode V - The Empire Strikes Back (1980)'' `Action', ``Mission: Impossible (1996)'' `Action', $\cdots$, ``Stargate (1994)'' `Action', ``\_'' `?', ``\_'' `?', $\cdots$, ``\_'' `?'
    & ``Star Wars: Episode V - The Empire Strikes Back (1980)'' `Action', ``Mission: Impossible (1996)'' `Action', $\cdots$, ``Stargate (1994)'' `\sethlcolor{yellow}\hl{Action}', ``?'' `\sethlcolor{yellow}\hl{Action}', ``?'' `\sethlcolor{yellow}\hl{Comedy}', $\cdots$, ``?'' `\sethlcolor{yellow}\hl{Comedy}' \\
    \hline

    \textbf{Output:} & \textbf{Output:} & \textbf{Output:} \\
    Action, Comedy, Drama
    & ``Star Wars: Episode V - The Empire Strikes Back (1980)'' `Action', ``Mission: Impossible (1996)'' `Action', $\cdots$, ``Stargate (1994)'' `Action', ``\_'' `Action', ``\_'' `Comedy', $\cdots$, ``\_'' `Comedy'
    & ``Star Wars: Episode V - The Empire Strikes Back (1980)'' `Action', ``Mission: Impossible (1996)'' `Action', $\cdots$, ``Stargate (1994)'' `Action', ``Saving Private Ryan (1998)'' `Action', ``Pretty Woman (1990)'' `Comedy', $\cdots$, ``Ghost (1990)'' `Comedy' \\
    \hline
    
  \end{tabular}}
\end{table*}

\section{Method}
In this section, we begin by introducing the fundamentals of utilizing LLMs in list-wise recommendation. We then provide a detailed description of our proposed DLCRec model, followed by an explanation of the data augmentation strategies employed.
\subsection{LLM for list-level recommendation}
This section outlines the fundamental concepts and methodologies for fine-tuning LLMs for specific recommendation tasks. 

\noindent\textbf{Instruction Tuning.}
Instruction Tuning plays a crucial role in enhancing LLMs' ability to follow human instructions~\cite{ouyang2022training}. In our work, we employ instruction tuning with Parameter-Efficient Fine-Tuning (PEFT)~\cite{he2021towards, lester2021power, houlsby2019parameter} for each sub-task to enable LLMs to perform specific tasks effectively.
For a given task $T$, we first organize our data into a training dataset $\mathcal{Z}=\{(x_i, y_i)\}_{i=1}^N$, where $x_i$ and $y_i$ represent the instruction and its corresponding response, respectively. We then fine-tune the model by optimizing the following autoregressive loss~\cite{brown2020language}:

\begin{equation}\label{eq:instruction_tuning}
\max_{\Phi} \sum_{(x,y) \in \mathcal{Z}} \sum_{t=1}^{|y|} \log P_{\Phi}(y_t | x, y_{<t})
\end{equation}
where $y_t$ is the $t$-th token of the $y$, $y_{<t}$ represents the tokens before $y_t$ and $\Phi$ is the original parameters of the LLM.

\noindent\textbf{Grounding strategies.}
In contrast to methods that limit LLMs to selecting items from a predefined candidate pool~\cite{liao2023llara,zhang2023collm}, DLCRec empowers LLMs to generate items across the entire item space. However, this increased flexibility introduces the risk of generating items not present in the original dataset. To mitigate this issue, we adopt a two-step grounding paradigm, similar to BIGRec~\cite{bao2023bi}.
In the first step, we fine-tune the LLM to adapt its outputs to specific recommendation scenarios, allowing it to generate items from the latent item representation space. Given the continuity of this space, the generated items may not correspond to real-world items. Therefore, in the second step, we map the embeddings generated by the fine-tuned LLM to their nearest neighbors within the real item embeddings encoded by the original LLM, ensuring that the model generates valid items from the dataset. For detailed grounding strategies, kindly refer to Appendix~\ref{appendix: ground}.

\subsection{DLCRec Framework}
To fully harness the potential of LLMs in controlling diversity within recommender systems, we propose a three-stage framework, termed DLCRec, which decomposes the list recommendation task into three distinct sub-tasks: Genre Prediction (GP), Genre Filling (GF), and Item Prediction (IP). As illustrated in Figure \ref{fig:framework}, DLCRec comprises two primary components: a training framework and a control framework.

In the training framework, we independently train the three sub-tasks using data derived from few-shot samples. This modular approach allows each sub-task to specialize in its respective objective, thereby enhancing the overall effectiveness of the recommendation process.
In the control framework, these sub-tasks are executed sequentially, with control numbers propagated through each stage to shape the final recommendation lists. Specifically, we first predict future genres based on control numbers, which then guide the subsequent genre-filling and item-predicting stages. This ensures that control numbers influence the entire recommendation process, enabling more precise diversity management in the recommendation lists.

In the following sections, we provide a detailed exploration of the training and control frameworks within DLCRec, emphasizing their key components and interactions.

\subsubsection{Training Framework}
For each of the three sub-tasks, we tailor the few-shot data into specialized training datasets. The specific task formats for each stage are detailed in Table \ref{tab:sub-tasks}. Similar to alpaca~\cite{alpaca}'s instruction-tuning format, each sample comprises three essential components:
\begin{itemize}[leftmargin=*]
\item \textbf{Instruction:} This defines the task's objective, providing context and guidance for the LLM to understand the expected outcome.
\item \textbf{Input:} This represents the user's interaction behaviors, such as historical preferences and interests.
\item \textbf{Output:} This serves as the label or target response that the LLM should generate, given the instruction and input.
\end{itemize}
To train the LLM, we combine the instruction and input into a prompt and then feed it into the model to generate the output. Below, we detail each sub-task and its specific function:

\begin{itemize}[leftmargin=*]
\item \textbf{Genre Prediction (GP):} LLMs are fine-tuned to predict future genres that may appear in the recommendation list, based on historical interaction trails (i.e., items and their corresponding genres). The output is a set of probable genres. The goal is to capture patterns and relationships between items and genres, enabling the model to make informed predictions about which genres are likely to appear in the future recommendation list. 

\item \textbf{Genre Filling (GF):} LLMs are instructed to fill in the future genre placeholders using the predicted genres while considering the historical interaction trails. This stage ensures that the model has a clear understanding of the genres that should be represented in the list, allowing it to generate items that align with these genres. 

\item \textbf{Item Prediction (IP):} LLMs are instructed to predict future items (i.e., filling the item placeholders) based on historical interaction trails and the predicted future genres. The filled genre placeholders from the Genre Filling task serve as guidance, enabling the model to generate items that achieve the desired genre coverage.
\end{itemize}

In tasks GF and IP, we employ a constrained format, where ``'' represents an item placeholder and `' represents a genre placeholder. The format mirrors the interaction trails, with `?' denoting tokens that need to be predicted and `\_' indicating irrelevant tokens that the model should disregard. This format not only enhances the performance of the tasks but also streamlines the transfer of information between them. In GF, the constrained format effectively aids LLMs in organizing genres, while in IP, it better guides LLMs in generating items that correspond to specific genres.

It is worth noting that we train different models for different sub-tasks, and adding data from other sub-tasks to the training set does not significantly improve performance.

\subsubsection{Control Framework}
After training the three sub-tasks, they can be integrated to address the diversity-oriented recommendation task by leveraging their specialized capabilities. The primary objective of DLCRec is to generate recommendation lists that meet varying diversity requirements, where users can specify how many or exactly which genres the lists should contain. 

To achieve this, it is crucial to control the input signals that target specific genre coverage in the recommendations. Below, we detail how each sub-task adjusts its prompt within the control framework:

\begin{itemize}[leftmargin=*]
\item \textbf{Genre Prediction (GP):} The output of task GP is controlled by restricting the number of target genres in the instruction prompt. By modifying this prompt and using the control number to specify the number of target genres, we can effectively guide the output of GP.

\item \textbf{Genre Filling (GF):} The output of task GF is managed by limiting the target genres in the instruction prompt, which dictates the genres used to fill the future genre tokens. Controlling these target genres, particularly by using the output from GP, allows us to shape the output of GF according to the control number.

\item \textbf{Item Prediction (IP):} The output of task IP is governed by modifying the future genres in the input prompt. By controlling these future genres, we can influence the genre composition of the recommendation lists. Specifically, the output from GF serves as the future genres, enabling the control number to direct IP's output effectively.

\end{itemize}

In summary, these three sub-tasks operate in a cascading manner, where the output of each sub-task influences the subsequent one. The control number propagates through the sequence, empowering the LLMs to regulate the diversity of the recommendations. This control framework allows DLCRec to generate recommendation lists with customizable diversity levels, catering to the specific needs of individual users.
 

\subsection{Data Augmentation}

\begin{figure}[!t]
    \centering
    \includegraphics[width=\linewidth]{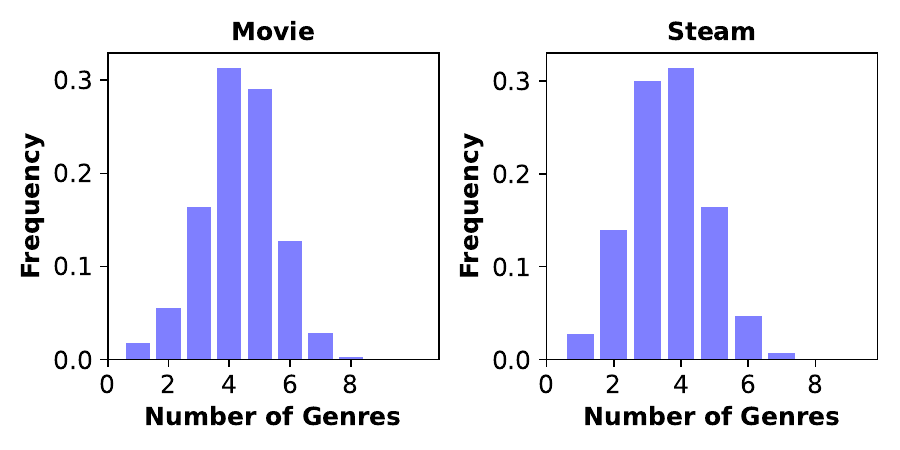}
    \caption{Distribution of the number of genres in the Movie and Steam datasets.}
    \label{fig:distribution}
\end{figure}

\begin{figure}[!t]
    \centering
    \includegraphics[width=0.98\linewidth]{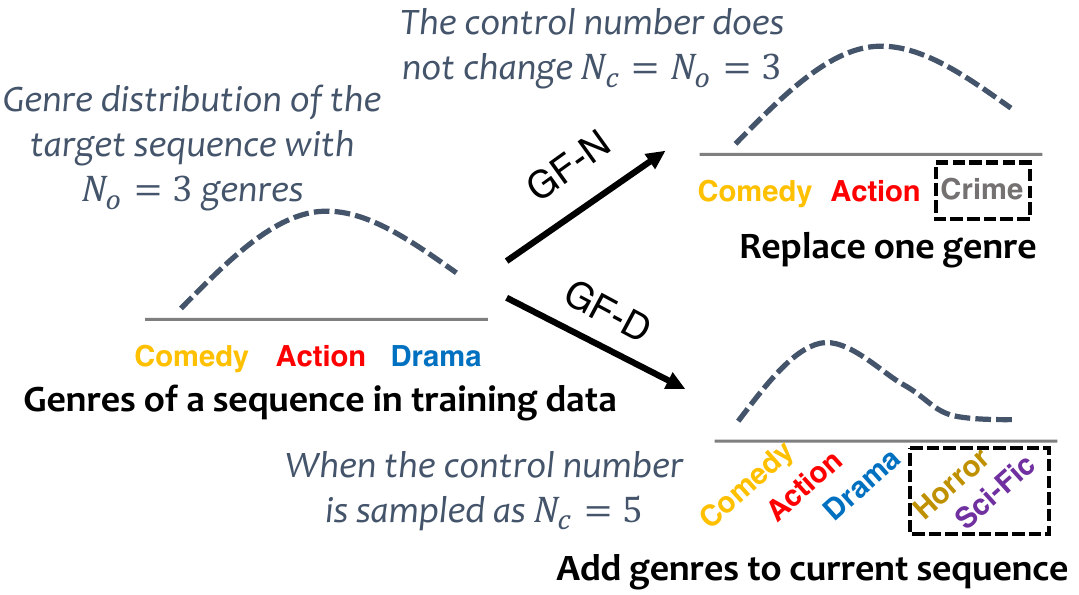}
    \caption{Illustration of the data augmentation method employed in task GF. Two strategies are applied: (a) ``GF-N'', which introduces noise by replacing the original genre with a noisy genre, and (b) ``GF-D'', which manipulates the genre distribution by randomly adding or deleting genres up to the sampled controlled threshold, promoting a uniform target distribution.}
    \label{fig:DA_GF}
\end{figure}

Generating recommendations with diverse levels of diversity is challenging due to the scarcity and skewness of diversity signals in the training data. As illustrated in Figure \ref{fig:distribution}, the distribution of the number of genres in the Movie~\cite{harper2015movielens} and Steam~\cite{kang2018self} datasets is uneven, peaking at 4 or 5 genres. This limitation can lead to degraded model performance when control targets are rare or unseen. To address this issue, we propose two novel data augmentation approaches for the GF and IP tasks, enhancing the model's robustness to noisy and out-of-distribution control targets. By combining the original data with the augmented data, we create a more diverse and representative training set, enabling the model to learn more generalizable patterns and improve its performance in generating recommendations with diverse levels of diversity.

\subsubsection{Data Augmentation in Task GF} 
As shown in Figure \ref{fig:DA_GF}, we employ two data augmentation strategies, GF-N and GF-D, to enhance the model's robustness to noise and varying control numbers in task GF.

\begin{itemize}[leftmargin=*]
\item \textbf{GF-N:} To increase the model's resilience to noisy target genres, we manually introduce noise into the training sequences by replacing one genre in each sequence with a noisy genre. This ensures that the number of controlled genres $N_c$ remains equal to the original genre number $N_o$.

\item \textbf{GF-D:} To make the model robust to varying control numbers $N_c$, we uniformly sample $N_c$ from the range of $(1,10)$ and adjust the genre interaction sequences accordingly. If the original number $N_o$ exceeds the control number $N_c$, we replace the least frequent genres with the remaining genres randomly. Conversely, if $N_o$ is smaller than $N_c$, we sample additional genres to reach $N_c$ based on the genre distribution in the training set.
\end{itemize}



\subsubsection{Data Augmentation in Task IP}
Following the GF-N and GP-D tasks, we further apply two data augmentation methods to the original item sequences in the training data: IP-N and IP-D.

\begin{itemize}[leftmargin=*]
\item \textbf{IP-N:} To account for potential errors in task GF's predictions, which may result from inaccuracies in both tasks GP and GF, we introduce noise into the input of the IP task. Specifically, we define an error rate $r$, representing the probability of incorrect predictions in the GF task. We then replace a proportion $r$ of the items in the recommendation list with items from other genres based on the item distribution in the training set.

\item \textbf{IP-D:} Following the GF-D strategy, we further sample items for each genre altered in GF-D. This ensures that the model is exposed to a wide range of items with varying genre coverage, enhancing its ability to generate items across different genres.
\end{itemize}

These data augmentation strategies equip our model to generate recommendations that meet both genre-level and item-level diversity requirements, thereby improving the model's robustness to noisy and out-of-distribution data.





\section{Experiments}
In this section, we conduct a comprehensive evaluation of our proposed DLCRec framework on two real-world datasets, benchmarking its performance against several baselines. Furthermore, we design two ablation studies to investigate the effectiveness of the task-decomposition framework and the data augmentation strategies in DLCRec. We also perform a sensitivity analysis to assess how well DLCRec adapts to varying control numbers. 

To thoroughly assess DLCRec's superiority in diversity control, we aim to answer the following research questions:

\begin{itemize}[leftmargin=*]
    \item \textbf{RQ1:} How does our method obey precise control numbers compared to other baseline methods?
    \item \textbf{RQ2:} How does the decomposition framework function in our method? Is it necessary?
    \item \textbf{RQ3:} How does the data augmentation method benefit our method?
\end{itemize}

\subsection{Experiments Setup}

\begin{table}[t]
\setlength{\abovecaptionskip}{-0.1cm}
\setlength{\belowcaptionskip}{-0.3cm}
\caption{Statistics of the two datasets.}
\begin{center}
\setlength{\tabcolsep}{2mm}
\resizebox{\columnwidth}{!}
{
\begin{tabular}{ccccc} 
\toprule
\textbf{Dataset} & \textbf{\# Item} & \textbf{\# Interaction} & \textbf{\# Sequence} & \textbf{\# Genre} \\ \midrule
Movie & 10,681 & 5,885,448 & 4,650,273 & 20 \\ 
Steam & 6,772 & 1,106,982 & 444,311 & 22 \\ \bottomrule
\end{tabular}
}
\end{center}
\label{tab:statistics}
\end{table}

\begin{table*}[!h]
\centering
\caption{Performance comparison of DLCRec and other LLM-based baselines on Movie and Steam datasets. The best results in control metrics are highlighted in bold.}
\label{tab:performance_comparison}
\begin{tabular}{l|cccc|cccc}
\toprule
\multirow{2}{*}{\textbf{Method}} & \multicolumn{4}{c|}{\textbf{Movie}} & \multicolumn{4}{c}{\textbf{Steam}}\\
& \textbf{NDCG@10} & \textbf{Recall@10} & \textbf{Cov@10} & \textbf{MAE\_Cov@10} & \textbf{NDCG@10} & \textbf{Recall@10} & \textbf{Cov@10} & \textbf{MAE\_Cov@10} \\
\midrule
\multicolumn{9}{c}{\textbf{control number=2}}\\
\midrule
\textbf{BIGRec\_div} & 0.0490 & 0.0428 & 2.526 & 1.112 & 0.0211 & 0.0173 & 2.689 & 0.827 \\
\textbf{BIGRec\_CoT} & 0.0466 & 0.0396 & 2.517 & 0.987 & 0.0268 & 0.0218 & 2.242 & 0.486 \\
\textbf{DLCRec} & 0.0357 & 0.0306 & 2.478 & \textbf{0.798} & 0.0236 & 0.0201 & 1.655 & \textbf{0.445} \\
\midrule
\multicolumn{9}{c}{\textbf{control number=5}}\\
\midrule
\textbf{BIGRec\_div} & 0.0481 & 0.0426 & 2.515 & 2.529 & 0.0218 & 0.0182 & 2.713 & 2.291 \\
\textbf{BIGRec\_CoT} & 0.0469 & 0.040 & 2.519 & 2.541 & 0.0277 & 0.0225 & 2.257 & 2.745 \\
\textbf{DLCRec} & 0.0450 & 0.0405 & 4.468 & \textbf{0.662} & 0.0289 & 0.0272 & 4.582 & \textbf{0.434} \\
\midrule
\multicolumn{9}{c}{\textbf{control number=8}}\\
\midrule
\textbf{BIGRec\_div} & 0.0480 & 0.0426 & 2.508 & 5.492 & 0.0216 & 0.0179 & 2.743 & 5.257 \\
\textbf{BIGRec\_CoT} & 0.0470 & 0.0403 & 2.525 & 5.475 & 0.0274 & 0.0223 & 2.253 & 5.747 \\
\textbf{DLCRec} & 0.0436 & 0.0383 & 7.495 & \textbf{0.511} & 0.0273 & 0.0248 & 7.873 & \textbf{0.131} \\
\bottomrule
\end{tabular}
\end{table*}

\subsubsection{Datasets.} 
We conduct experiments on two real-world datasets, with statistics summarized in Table \ref{tab:statistics}. 

\begin{itemize}[leftmargin=*]
\item \textbf{Movie~\cite{harper2015movielens}}. This is a widely-used dataset for movie recommendations — MovieLens10M\footnote{https://grouplens.org/datasets/movielens/10m/}.
\item \textbf{Steam~\cite{kang2018self}}. This dataset contains user-game interaction data from the Steam store, a popular online gaming platform.
\end{itemize}

For both datasets, we focus on positive interactions, defined as ratings above 3 for MovieLens10M and playing time exceeding 3 hours for Steam. To facilitate model training, we set the sequence length to 20, using the first 10 interactions as historical sequences and the latter 10 as future sequences. We discard sequences with fewer than 20 interactions, sort the remaining ones chronologically, and split them into training, validation, and testing sets at a ratio of 8:1:1. We then randomly sample 1,000 sequences from each dataset to form the final training, validation, and testing sets. To simplify diversity control, we only consider the primary genre of each item.

\subsubsection{Evaluation.}
To comprehensively evaluate the performance of DLCRec and other baselines in generating recommendations with varying levels of diversity, we design a multi-faceted evaluation framework with three distinct control settings, each with a different control number: Low (2 genres), Medium (5 genres), and High (8 genres). These control numbers are strategically chosen to represent both rare and common control requirements in the original dataset, allowing us to assess the models' ability to adapt to diverse user preferences. We employ two accuracy metrics: NDCG@K\cite{jarvelin2002cumulated} and Recall@K, to assess the relevance of the recommended items. Additionally, we use two control metrics: Cov@K and MAE\_Cov@K, to evaluate the model's ability to meet the diversity requirements. Cov@K measures the coverage of genres in the recommended list, while MAE\_Cov@K calculates the mean average error between the actual and desired genre coverage. Our goal is to achieve a lower MAE\_Cov@K while maintaining high accuracy, indicating that the model can effectively balance relevance and diversity.

\subsubsection{Implementing Details.}
We employ Llama3\cite{dubey2024llama} as the backbone LLM for our experiments. For all LLM-based methods, we adopt a unified training protocol: each model is trained for at most 25 epochs with a batch size of 250. The learning rate is tuned within the range of [1e-4, 5e-4, 1e-3] to optimize performance. We utilize the AdamW optimizer\cite{loshchilov2017decoupled} for model training. During inference, we employ greedy search to generate the output sequences. We set the error rate $r$ in the IP-N strategy as 0.3.

\subsection{Overall Performances (RQ1)}
In this section, we compare DLCRec with controllable LLM-based methods in the settings of different control numbers. 

\subsubsection{Baselines.}
We identify two classical approaches for diversity control based on existing literature~\cite{liang2024taxonomy,gao2024llm,wei2022chain}. Although these methods were originally developed for different scenarios, we adapt and simplify them for our specific context, categorizing them into two main baselines:
\begin{itemize}[leftmargin=*]

\item \textbf{BIGRec\_div~\cite{bao2023bi}}. We adapt BIGRec to accommodate diverse recommendations by fine-tuning it with prompts that incorporate different control numbers. Since BIGRec is originally designed as a sequential recommender for single-item prediction, we modify the task objective to enable the model to generate a list of recommendations instead.
\item \textbf{BIGRec\_CoT~\cite{wei2022chain}}. We integrate the chain-of-thought prompting strategy with BIGRec to facilitate diverse recommendations. Specifically, we first prompt the model to generate genres corresponding to the control numbers, and then ask it to produce recommendations based on the inferred genres. This approach enables the model to explicitly consider the desired diversity in the recommendation process.
\end{itemize}

\subsubsection{Results.} 
As shown in Table \ref{tab:performance_comparison}, DLCRec achieves the lowest MAE\_Cov@10 under all circumstances, indicating that it most successfully adheres to the control numbers for diversity in the recommendation lists, thereby demonstrating its effectiveness in meeting the diversity requirements. In contrast, BIGRec\_div and BIGRec\_CoT, despite modifying prompts to instruct LLMs to recommend based on control numbers through instruction tuning, struggle with mastering this complex task. DLCRec's approach of decomposing the fine-tuning process enables it to excel in this diversity-oriented controllable recommendation task.

Regarding accuracy, we observe that DLCRec makes a minor trade-off in accuracy to gain controllability. However, this trade-off is relatively small, particularly when the control numbers are 5 and 8. Notably, DLCRec even surpasses the other two baselines in the Steam dataset when the control number is 5, achieving a lower MAE\_Cov while maintaining competitive accuracy. This showcases DLCRec's ability to balance accuracy and controllability in recommendation tasks.

\begin{figure}[tb!]
\setlength{\abovecaptionskip}{-0cm}
\setlength{\belowcaptionskip}{-0cm}
\centering
\includegraphics[scale=0.55]{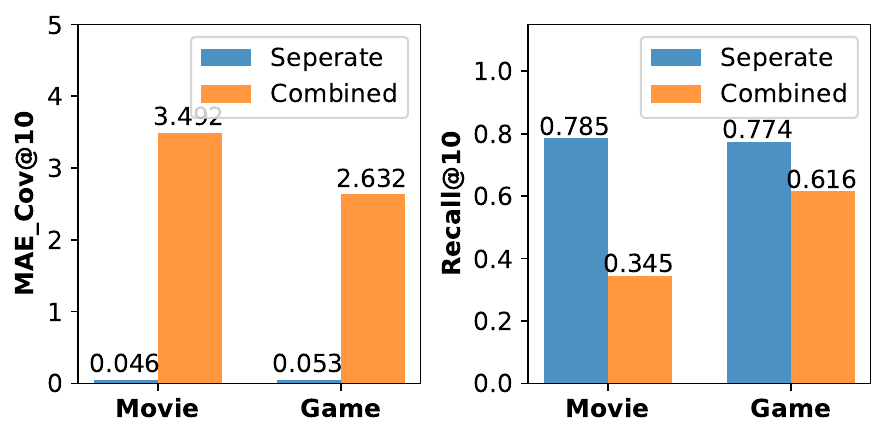}
\caption{Performance comparison of separating and combining tasks GP and GF.}
\label{fig:abs_framework}
\end{figure}

\begin{table*}[h]
\centering
\tabcolsep=2pt
\caption{Comparison of performances of DLCRec with different data augmentation strategies. "DA" denotes "Data Augmentation". "GF-ND" and "IP-ND" represent models using both data augmentation techniques, while "GF" and "IP" represent models without data augmentation. The best results in each group are highlighted in bold.}
\label{tab:da}
\begin{tabular}{cc|cccc|cccc}
\noalign{\vskip 1pt}
\toprule
\noalign{\vskip 1pt}
\multirow{2}{*}{\textbf{IP DA}} & \multirow{2}{*}{\textbf{GF DA}} & \multicolumn{4}{c|}{\textbf{Movie}} & \multicolumn{4}{c}{\textbf{Steam}} \\

 &  & \textbf{NDCG@10} & \textbf{Recall@10} & \textbf{COV@10} & \textbf{MAE\_Cov@10} & \textbf{NDCG@10} & \textbf{Recall@10} & \textbf{COV@10} & \textbf{MAE\_Cov@10} \\
\midrule
\textbf{IP} & \textbf{GF} & 0.0390 & 0.0354 & 3.925 & 1.133 & 0.0274 & 0.0256 & 4.300 & 0.704 \\
\midrule
\textbf{IP-ND} & \textbf{GF} & 0.0430 & 0.0396 & 4.426 & 0.710 & 0.0265 & 0.0249 & 4.529 & 0.485 \\
\textbf{IP-ND} & \textbf{GF-N} & 0.0437 & 0.0399 & 4.509 & \textbf{0.603} & 0.028 & 0.0262 & 4.524 & 0.494 \\
\textbf{IP-ND} & \textbf{GF-D} & 0.0425 & 0.0381 & 4.487 & 0.611 & 0.0260 & 0.0242 & 4.822 & \textbf{0.192} \\
\textbf{IP-ND} & \textbf{GF-ND} & \textbf{0.0450} & \textbf{0.0405} & 4.468 & 0.662 & \textbf{0.0289} & \textbf{0.0272} & 4.582 & 0.434 \\
\midrule
\textbf{IP} & \textbf{GF-ND} & 0.0449 & \textbf{0.0416} & 3.927 & 1.109 & 0.0281 & 0.026 & 4.338 & 0.666 \\
\textbf{IP-N} & \textbf{GF-ND} & 0.0431 & 0.0390 & 4.296 & 0.740 & 0.0240 & 0.0214 & 4.514 & 0.492 \\
\textbf{IP-D} & \textbf{GF-ND} & \textbf{0.0454} & 0.0395 & 3.845 & 1.251 & 0.0282 & 0.0254 & 4.799 & \textbf{0.219} \\
\textbf{IP-ND} & \textbf{GF-ND} & 0.0450 & 0.0405 & 4.468 & \textbf{0.662} & \textbf{0.0289} & \textbf{0.0272} & 4.582 & 0.434 \\
\noalign{\vskip 1pt}
\bottomrule
\end{tabular}
\end{table*}

\subsection{Ablation Study (RQ2 \& RQ3)}
In this section, we conduct ablation studies to validate the importance of DLCRec's task decomposition and data augmentation. 

\subsubsection{Decomposition framework.} We conduct experiments to testify our decomposition framework is necessary. As shown in Table \ref{tab:performance_comparison}, using a single task to fine-tune LLMs fails to enable them to adapt recommendations to different diversity requirements. Moreover, our experiments reveal that combining tasks GP and GF is not a viable solution. As illustrated in Figure \ref{fig:abs_framework}, combining tasks GP and GF results in significantly lower performance on future genre filling, as measured by the control metric MAE\_Cov@10 and the genre accuracy metric Recall@10. Notably, when tasks GP and GF are combined, the output is dominated by the most popular genre in history, indicating that the model is unable to learn the genre-filling task and instead produces homogeneous results. This highlights the importance of DLCRec's decomposition framework in enabling effective diversity-oriented controllable recommendations.

\subsubsection{Data Augmentation.} To investigate the impact of data augmentation on the performance of DLCRec in tasks GF and GP, we conducted an ablation study by experimenting with different combinations of data augmentation strategies. The results are presented in Table \ref{tab:da}, which shows the performance of DLCRec with various data augmentation strategies. To isolate the effect of each strategy, we kept one task stable while analyzing the performance trends of the other. For brevity, we only report the results for a control number of 5; additional results can be found in the Appendix \ref{appendix: abs}.

Our analysis yields the following key findings:
\begin{itemize}[leftmargin=*]
\item The combinations of "GF-ND" and "IP-ND", which employ both data augmentation strategies in tasks GF and IP, achieve the best overall performance, with high accuracy and precise control across both datasets. This suggests that these combinations strike the optimal balance between accuracy and diversity control.
\item Strategy "GF-D" and "IP-D", which involve adding data with diverse control numbers, demonstrate a strong ability to reduce coverage control errors, particularly in the Steam dataset.
\item Strategy "GF-N" and "IP-N", which introduce noisy inputs, also yield improvements in control abilities. However, employing "IP-D" alone may compromise accuracy, as the induced noise can interfere with the LLM's item prediction capability during training. Notably, when both data augmentation strategies are combined in task IP, they complement each other, resulting in a more robust model. 
\end{itemize}

These findings highlight the importance of data augmentation strategies in enhancing the performance of DLCRec and demonstrate the value of carefully selecting the most effective combinations of strategies to achieve optimal results.

\begin{figure}[tb!]
\setlength{\abovecaptionskip}{-0cm}
\setlength{\belowcaptionskip}{-0.4cm}
\centering
\includegraphics[scale=0.38]{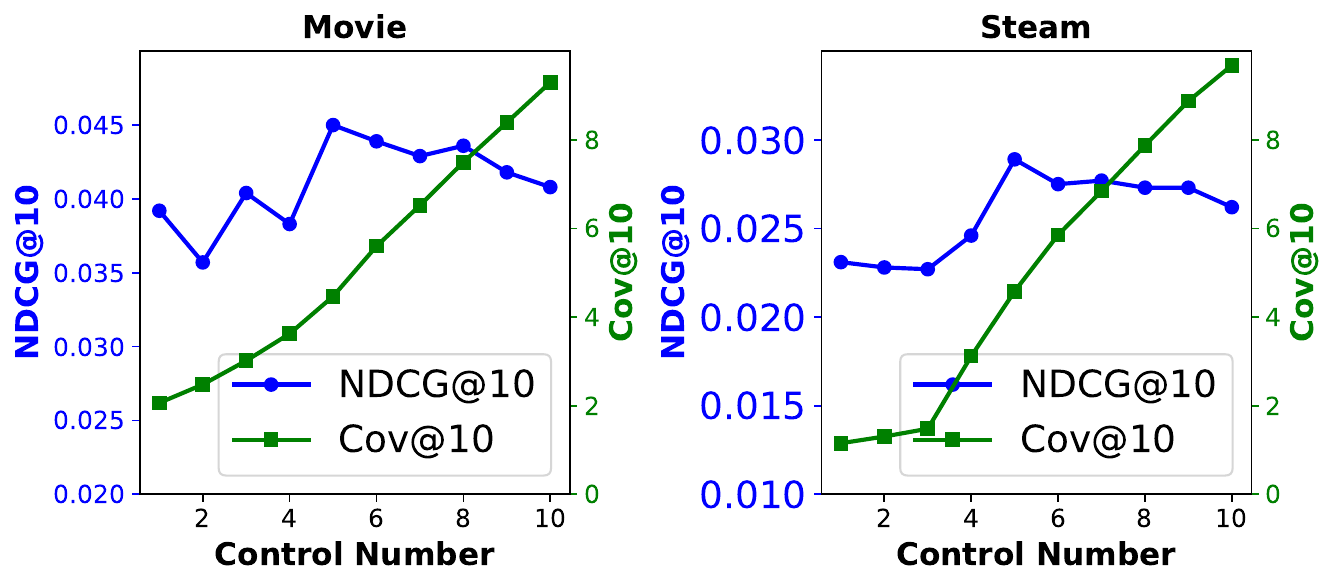}
\caption{Sensitivity of DLCRec to varying levels of diversity}
\label{fig:1to10}
\end{figure}

\subsection{Sensitivity Analysis} \label{Sensitivity}
To further evaluate DLCRec's control capacity, we experiment to test its ability to produce recommendations with varying levels of diversity. Specifically, we manipulated the control number to range from 1 to 10, corresponding to the minimum and maximum diversity, respectively. This setup enabled us to determine whether DLCRec could accurately produce recommendations that align with the user's desired diversity level.

The results, depicted in Figure \ref{fig:1to10}, demonstrate that DLCRec's recommendations exhibit a strong correlation with the control number. As the control number increases from 1 to 10, the coverage of genres in the recommended items also increases in a nearly linear fashion, with a slope slightly less than 1. This minor deviation from a perfect linear relationship can be attributed to the inherent challenges in achieving complete genre alignment due to the limitations of LLMs' genre mapping capabilities and the grounding mechanism.

Notably, the accuracy of DLCRec's recommendations remains relatively stable across the range of control numbers, indicating that the model can adapt to varying diversity requirements without compromising its accuracy. In both datasets, the highest accuracy is achieved when the control number is 5, which is likely due to the prevalence of this signal in the training distribution. In the Steam dataset, we observe that DLCRec tends to generate recommendations with a smaller coverage when the control number is small, which can be attributed to the Matthew effect~\cite{gao2023alleviating} caused by the dominant "Action" genre in the Steam dataset. Overall, DLCRec exhibits the ability to flexibly cater to users' diverse needs while maintaining a consistent level of recommendation quality.

\section{Conclusion}
In this work, we introduce DLCRec, a novel diversity-oriented controllable recommendation framework that decomposes the task into three sub-tasks: genre predicting, genre filling, and item predicting. These sub-tasks are trained independently and inferred sequentially according to user-defined control numbers, ensuring more precise control over diversity. 
To address the challenges of scarcity and skewness of training data, we employ two data augmentation strategies to enhance the model's robustness and adaptability to different diversity requirements. Through extensive experiments, we demonstrate the effectiveness of DLCRec in adjusting recommendation diversity and validate the decomposition framework and data augmentation strategies.

This work presents a promising approach to fine-tuning LLMs for controllable recommendations. While our primary focus is on the diversity, the proposed framework can be adapted to more fine-grained tasks. Future research directions could include optimizing the generation capabilities of LLMs for list-level recommendations, further improving the flexibility and effectiveness of controllable recommendation systems. 

\section{Acknowledgements}
This work is supported by the National Natural Science Foundation of China (No.62402470, No.62272437, No.62121002, No.U21B2026), Anhui Provincial Natural Science Foundation (2408085QF189), the Fundamental Research Funds for the Central Universities of China (WK2100000053, PA2024GDSK0107), and the Postdoctoral Fellowship Program of CPSF (GZC20241643). This research is funded by KuaiShou and supported by the advanced computing resources provided by the Supercomputing Center of the USTC.

\bibliographystyle{ACM-Reference-Format}
\balance
\bibliography{DLCRec}

\newpage
\appendix

\section{Implementation Details}
\subsection{DLCRec Grounding Strategies}
\label{appendix: ground}

We implement a two-step grounding strategy to ensure the model generates valid items in the IP sub-task.

First, we fine-tune the large language model (LLM) on the training set to generate item lists. To clearly delineate individual items in the output, we enclose each item with special tokens ``''.

Second, we utilize the original, frozen LLM to encode both generated and valid items. Specifically, we extract embeddings from the last hidden layer's final token for each item. We then map each generated item to its nearest valid item based on L2 distance between their embeddings.

This grounding mechanism ensures the model outputs only valid items during inference, enhancing the overall accuracy and reliability. We apply a similar approach to baseline models like BIGRec~\cite{bao2023bi}, with the only difference being the use of alternative special tokens for item separation.

\subsection{Further Discussions}
\label{appendix: discussion}
Since sub-tasks GF and IP share similar input formats, we initially hypothesized that training a unified model would lead to a performance boost. However, we did not observe any significant improvement in performance for any combination of sub-tasks. As a result, in DLCRec, we opt to train separate models for each sub-task, utilizing only its specific training data along with its corresponding augmented data. We carried out all the fine-tuning tasks on an 80GB A100.

\section{Full data of ablation study of data augmentation}
\label{appendix: abs}
To further validate the effectiveness of different combinations of data augmentation strategies, we conducted experiments testing DLCRec's performance in precise control with varying control numbers. In addition to the control number 5 discussed in the main paper, we present the results for control numbers 2 and 8 in Tables \ref{tab:da control2} and \ref{tab:da control8}, respectively.

For control number 2:
Our data augmentation strategies show some improvement in MAE\_Cov@10 for the Movie dataset. However, they do not perform as well in the Steam dataset. We attribute this discrepancy to the dominance of the "Action" genre in the Steam dataset, which initially encourages the model to output less diversified recommendations and hinders its capacity to learn from a limited number of genres. Notably, performance improves significantly with larger control numbers (5 and 8), demonstrating our model's robustness to datasets with unevenly distributed genres.

For control number 8:
The combinations "GF-ND" and "IP-ND" achieve the best overall performance, exhibiting high accuracy and low error rates in coverage control across both datasets. This success can be attributed to the strong robustness of all the data augmentation strategies when applied to a larger control number.

These findings highlight our model's adaptability to different control numbers and dataset characteristics, reinforcing the effectiveness of our proposed data augmentation strategies in diversity-oriented controllable recommendations.

\begin{table*}[t]
\centering
\tabcolsep=2pt
\caption{Comparison of performances of DLCRec with different data augmentation strategies when the control number is 2. "DA" denotes "Data Augmentation". "GF-ND" and "IP-ND" represent models using both data augmentation techniques, while "GF" and "IP" represent models without data augmentation. The best results in each group are highlighted in bold.}
\label{tab:da control2}
\begin{tabular}{cc|cccc|cccc}
\noalign{\vskip 1pt}
\toprule
\noalign{\vskip 1pt}
\multirow{2}{*}{\textbf{IP DA}} & \multirow{2}{*}{\textbf{GF DA}} & \multicolumn{4}{c|}{\textbf{Movie}} & \multicolumn{4}{c}{\textbf{Steam}} \\

 &  & \textbf{NDCG@10} & \textbf{Recall@10} & \textbf{COV@10} & \textbf{MAE\_Cov@10} & \textbf{NDCG@10} & \textbf{Recall@10} & \textbf{COV@10} & \textbf{MAE\_Cov@10} \\
\midrule
\textbf{IP} & \textbf{GF} & 0.036 & 0.0321 & 2.479 & 0.843 & 0.0238 & 0.0202 & 1.651 & 0.449 \\
\midrule
\textbf{IP-ND} & \textbf{GF} & 0.0349 & 0.0301 & 2.448 & 0.788 & 0.0227 & 0.0194 & 1.301 & 0.809 \\
\textbf{IP-ND} & \textbf{GF-N} & 0.0356 & 0.0307 & 2.463 & \textbf{0.783} & 0.0226 & 0.0194 & 1.299 & 0.809 \\
\textbf{IP-ND} & \textbf{GF-D} & \textbf{0.0365} & \textbf{0.0315} & 2.677 & 0.877 & 0.0227 & 0.0194 & 1.3 & \textbf{0.806} \\
\textbf{IP-ND} & \textbf{GF-ND} & 0.0357 & 0.0306 & 2.478 & 0.798 & \textbf{0.0228} & \textbf{0.0195} & 1.304 & 0.81 \\
\midrule
\textbf{IP} & \textbf{GF-ND} & 0.035 & 0.0308 & 2.511 & 0.849 & 0.0236 & 0.0201 & 1.655 & 0.445 \\
\textbf{IP-N} & \textbf{GF-ND} & 0.0339 & 0.0296 & 2.285 & \textbf{0.709} & 0.0233 & \textbf{0.0206} & 1.483 & \textbf{0.623} \\
\textbf{IP-D} & \textbf{GF-ND} & \textbf{0.039} & \textbf{0.0337} & 2.402 & 0.81 & \textbf{0.0246} & 0.0198 & 1.495 & 0.649 \\
\textbf{IP-ND} & \textbf{GF-ND} & 0.0357 & 0.0306 & 2.478 & 0.798 & 0.0228 & 0.0195 & 1.304 & 0.81 \\
\noalign{\vskip 1pt}
\bottomrule
\end{tabular}
\end{table*}

\begin{table*}[t!]
\centering
\tabcolsep=2pt
\caption{Comparison of performances of DLCRec with different data augmentation strategies when the control number is 5. "DA" denotes "Data Augmentation". "GF-ND" and "IP-ND" represent models using both data augmentation techniques, while "GF" and "IP" represent models without data augmentation. The best results in each group are highlighted in bold.}
\label{tab:da control5}
\begin{tabular}{cc|cccc|cccc}
\noalign{\vskip 1pt}
\toprule
\noalign{\vskip 1pt}
\multirow{2}{*}{\textbf{IP DA}} & \multirow{2}{*}{\textbf{GF DA}} & \multicolumn{4}{c|}{\textbf{Movie}} & \multicolumn{4}{c}{\textbf{Steam}} \\

 &  & \textbf{NDCG@10} & \textbf{Recall@10} & \textbf{COV@10} & \textbf{MAE\_Cov@10} & \textbf{NDCG@10} & \textbf{Recall@10} & \textbf{COV@10} & \textbf{MAE\_Cov@10} \\
\midrule
\textbf{IP} & \textbf{GF} & 0.0390 & 0.0354 & 3.925 & 1.133 & 0.0274 & 0.0256 & 4.300 & 0.704 \\
\midrule
\textbf{IP-ND} & \textbf{GF} & 0.0430 & 0.0396 & 4.426 & 0.710 & 0.0265 & 0.0249 & 4.529 & 0.485 \\
\textbf{IP-ND} & \textbf{GF-N} & 0.0437 & 0.0399 & 4.509 & \textbf{0.603} & 0.028 & 0.0262 & 4.524 & 0.494 \\
\textbf{IP-ND} & \textbf{GF-D} & 0.0425 & 0.0381 & 4.487 & 0.611 & 0.0260 & 0.0242 & 4.822 & \textbf{0.192} \\
\textbf{IP-ND} & \textbf{GF-ND} & \textbf{0.0450} & \textbf{0.0405} & 4.468 & 0.662 & \textbf{0.0289} & \textbf{0.0272} & 4.582 & 0.434 \\
\midrule
\textbf{IP} & \textbf{GF-ND} & 0.0449 & \textbf{0.0416} & 3.927 & 1.109 & 0.0281 & 0.026 & 4.338 & 0.666 \\
\textbf{IP-N} & \textbf{GF-ND} & 0.0431 & 0.0390 & 4.296 & 0.740 & 0.0240 & 0.0214 & 4.514 & 0.492 \\
\textbf{IP-D} & \textbf{GF-ND} & \textbf{0.0454} & 0.0395 & 3.845 & 1.251 & 0.0282 & 0.0254 & 4.799 & \textbf{0.219} \\
\textbf{IP-ND} & \textbf{GF-ND} & 0.0450 & 0.0405 & 4.468 & \textbf{0.662} & \textbf{0.0289} & \textbf{0.0272} & 4.582 & 0.434 \\
\noalign{\vskip 1pt}
\bottomrule
\end{tabular}
\end{table*}

\begin{table*}[t!]
\centering
\tabcolsep=2pt
\caption{Comparison of performances of DLCRec with different data augmentation strategies when the control number is 8. "DA" denotes "Data Augmentation". "GF-ND" and "IP-ND" represent models using both data augmentation techniques, while "GF" and "IP" represent models without data augmentation. The best results in each group are highlighted in bold.}
\label{tab:da control8}
\begin{tabular}{cc|cccc|cccc}
\noalign{\vskip 1pt}
\toprule
\noalign{\vskip 1pt}
\multirow{2}{*}{\textbf{IP DA}} & \multirow{2}{*}{\textbf{GF DA}} & \multicolumn{4}{c|}{\textbf{Movie}} & \multicolumn{4}{c}{\textbf{Steam}} \\

 &  & \textbf{NDCG@10} & \textbf{Recall@10} & \textbf{COV@10} & \textbf{MAE\_Cov@10} & \textbf{NDCG@10} & \textbf{Recall@10} & \textbf{COV@10} & \textbf{MAE\_Cov@10} \\
\midrule
\textbf{IP} & \textbf{GF} & 0.0402 & 0.0355 & 4.699 & 3.301 & 0.0267 & 0.0245 & 6.707 & 1.293 \\
\midrule
\textbf{IP-ND} & \textbf{GF} & 0.043 & \textbf{0.0396} & 6.186 & 1.818 & 0.0272 & \textbf{0.025} & 7.716 & 0.288 \\
\textbf{IP-ND} & \textbf{GF-N} & 0.0418 & 0.0375 & 7.024 & 0.98 & \textbf{0.0275} & {0.0249} & 7.745 & 0.261 \\
\textbf{IP-ND} & \textbf{GF-D} & 0.0431 & 0.038 & 7.49 & 0.516 & \textbf{0.0275} & {0.0249} & 7.871 & 0.133 \\
\textbf{IP-ND} & \textbf{GF-ND} & \textbf{0.0436} & 0.0383 & 7.495 & \textbf{0.511} & 0.0273 & 0.0248 & 7.873 & \textbf{0.131} \\
\midrule
\textbf{IP} & \textbf{GF-ND} & 0.0437 & \textbf{0.0392} & 5.547 & 2.455 & \textbf{0.0275} & \textbf{0.0251} & 6.88 & 1.12 \\
\textbf{IP-N} & \textbf{GF-ND} & 0.0405 & 0.0348 & 7.154 & 0.848 & 0.0217 & 0.0182 & 7.759 & 0.247 \\
\textbf{IP-D} & \textbf{GF-ND} & \textbf{0.0439} & 0.0373 & 7.077 & 0.927 & 0.0231 & 0.0206 & 7.908 & \textbf{0.092} \\
\textbf{IP-ND} & \textbf{GF-ND} & 0.0436 & 0.0383 & 7.495 & \textbf{0.511} & 0.0273 & 0.0248 & 7.873 & 0.131 \\
\noalign{\vskip 1pt}
\bottomrule
\end{tabular}
\end{table*}
\end{document}